\documentclass[aps,pra,twocolumn,superscriptaddress,nopacs,letterpaper,longbibliography]{revtex4-2}

\usepackage[utf8]{inputenc}
\usepackage{amsmath,stmaryrd}
\usepackage{color,graphicx}
\usepackage[pdfstartview=FitH,colorlinks=true,linkcolor=blue,citecolor=blue,urlcolor=blue]{hyperref}
\newcommand{\1}{\uparrow}
\newcommand{\2}{\downarrow}
\usepackage{lmodern}
\usepackage{orcidlink}
\usepackage{times}

\begin{document}

\title{Josephson dynamics and localization revivals in ultradilute quantum liquids}

\author{Piotr Wysocki\orcidlink{0009-0009-1589-1524}}
\affiliation{Faculty of Physics, Warsaw University of Technology, Ulica Koszykowa 75, PL-00662 Warsaw, Poland}
\affiliation{Faculty of Physics, University of Warsaw, Pasteura 5, PL-02093 Warsaw, Poland}
\author{Krzysztof Jachymski\orcidlink{0000-0002-9080-0989}}
\email{krzysztof.jachymski@fuw.edu.pl}
\affiliation{Faculty of Physics, University of Warsaw, Pasteura 5, PL-02093 Warsaw, Poland}
\author{Grigori E. Astrakharchik\orcidlink{0000-0003-0394-8094}}
\affiliation{Departament de Física, Universitat Politècnica de Catalunya, E-08034 Barcelona, Spain}
\author{Marek Tylutki\orcidlink{0000-0002-4243-3803}}
\email{marek.tylutki@pw.edu.pl}
\affiliation{Faculty of Physics, Warsaw University of Technology, Ulica Koszykowa 75, PL-00662 Warsaw, Poland}

\date{\today}

\begin{abstract}
We study the Josephson junction made of a one-dimensional ultradilute quantum liquid in a double-well potential. We analyze the dynamics as a function of the interaction strength and compare the results to the standard bosonic Josephson junction. It is found that the beyond-mean-field effects alter the dynamics, particularly in the regime, where the beyond-mean-field corrections dominate over the residual mean-field interaction. In that case, we observe nonlinear dynamics describing localization revivals instead of regular Josephson oscillations. In the regime where the ultradilute quantum liquids perform the regular Josephson oscillations, their frequency is also significantly changed in comparison with the regular Josephson junction. These results provide experimental characteristics of the ultradilute quantum liquids that contrast with the Josephson oscillations of a regular Bose-Einstein condensate. 
\end{abstract}

\maketitle

\section{Introduction}
Ultracold neutral atoms have proven to be a perfect setup for studying fundamental quantum phenomena as well as a promising platform for near-term quantum devices~\cite{Schafer2020,Amico2021}. Low temperatures make quantum effects more prominent by reducing thermal fluctuations and extending the system's coherence time. A prominent and paradigmatic example is the creation of a Bose-Einstein condensate (BEC) in ultracold atomic gases~\cite{KetterleN,CornellN}. Quantum degenerate bosonic gases allow for a remarkably simple description in terms of a macroscopic wave function which follows the nonlinear Gross-Pitaevskii equation (GPE)~\cite{Dalfovo1999, Pitaevskii2016}. However, the properties of the system can be far from trivial. For instance, in binary mixtures with balanced repulsive and attractive interactions near the border of stability the beyond-mean-field (BMF) effects described by the Lee-Huang-Yang (LHY) term~\cite{LHY,HP} become comparable in strength with the mean-field interaction. As a result, the formation of a novel state of matter corresponding to an ultradilute quantum liquid rather than a gas was predicted~\cite{Petrov2015}. Such states differ from the usual liquids in two key aspects, as quantum liquids (i) possess equilibrium density which is, by many orders of magnitude, more dilute than that of typical liquids, and (ii) are fully coherent, meaning they are still described by a single condensate function. This latter feature makes quantum liquids qualitatively different, as they possess a well-defined phase that is absent in classical liquids. This phase coherence can potentially manifest itself in phase-related properties such as interference and the Josephson effect.

In finite-size systems, there is an interplay between bulk and surface energies, leading to a formation of quantum droplets, which can be formed in one-, two- and three-dimensional geometries~\cite{Petrov2015,Petrov2016,Ilg2018,Zin2018}. Experimental realization of quantum droplets in binary mixtures was achieved shortly after their theoretical proposal~\cite{Cabrera2018,Cheiney2018,Semeghini2018,Ferioli2019,DErrico2019}. Due to a near cancellation of the repulsion and attraction, a separation of scales exists, describing soft and hard modes. As a result, the energy contribution that is relevant for short scales can be incorporated as a local term in the Gross-Pitaevskii equation for the liquid. Furthermore, within the GPE approach, the interaction regime of a one-dimensional (1D) quantum droplet can also be parametrized by a single control parameter, i.e. the rescaled number of particles and its excitation spectrum and dynamics has nontrivial properties, as studied in Refs.~\cite{Astrakharchik2018,Tylutki2020}. Quantum droplets can also arise in bosonic systems with dipolar interactions, close to the regime in which contact repulsion is balanced by dipolar attraction, as studied extensively in recent years both theoretically and experimentally~\cite{Kadau2016,FerrierBarbut2016,Chomaz2016,Wachtler2016,Wachtler2016a,Bisset2016,FerrierBarbut2018}. 
\begin{figure}
\includegraphics[clip = true, width = .99\columnwidth]{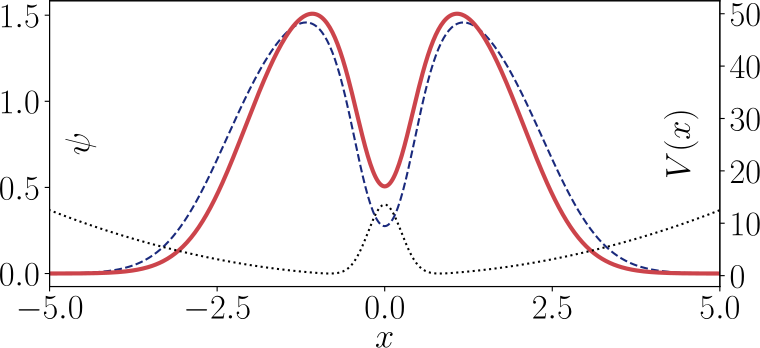}\\
\caption{Characteristic examples of the condensate function $\Psi(x)$ of a system in a double-well potential considering the case with $\delta g > 0$. 
The dashed line represents the gas described solely by the mean-field terms, given by Eq.~(\ref{eq.gpe0}), while the solid line includes the beyond-mean-field (BMF) correction, given by Eq.~(\ref{eq.gpe}). Although the addition of the Lee-Huang-Yang (LHY) term leads only to a slight change in the density profile, it significantly affects the dynamics within the double well. The dotted line shows the potential (scale on the right)}
\label{fig.wavef}
\end{figure}
Studying the stationary states of quantum systems provides important insight into their properties. However, examining their dynamics allows for a deeper understanding of not only small amplitude excitations, but also the rich landscape of out-of-equilibrium phenomena, such as dynamical phase transitions~\cite{Heyl2018}. One of the fundamental examples, characteristic of quantum superfluid systems, is the Josephson junction (JJ). The dynamics of Josephson junctions has been extensively studied in the context of ultracold atomic gases both in theory~\cite{Javanainen1986,Smerzi1997,Meier2001} and in experiment~\cite{Albiez2005,Valtolina2015,Spagnolli2017,Zhang2024}. Josephson oscillations can occur between two weakly coupled superfluid systems. A simple implementation requires a double-well external trapping potential which creates two superfluid pools with weak tunneling between them. Such systems are easily achievable in experiments with ultracold gases due to the precise control over the interactions and external potentials. A small imbalance in the population of the two wells leads to oscillations in both particle number and relative phase with a frequency much smaller than the trap frequency, which is closely related to the excitation spectrum of the system. However, for a large initial imbalance, the system enters a nonlinear dynamical regime known as a self-trapping characterized by small oscillations around a nonzero average imbalance. The mechanism behind oscillatory dynamics in JJs can be understood in terms of the two-mode model (TMM) which simplifies the description of the system to considering only two spatial modes corresponding to occupations of either of the wells~\cite{Smerzi1997}. Its validity has been the subject of detailed studies~\cite{Ananikian2006,JuliaDiaz2010}. A more detailed description of Josephson oscillations can be inferred from the Bogoliubov excitation spectrum, where the Josephson oscillations are related to the first antisymmetric mode~\cite{Morgan1998,Burchianti2017,MartinezGaraot2018}. In addition, there were numerous numerical studies of the Josephson dynamics and the dynamical phase diagram both in bosonic and in fermionic systems~\cite{Jezek2013,Xhani2020,Wlazlowski2023}. 

A renewed interest in Josephson dynamics has been brought by the recent experimental realization of the self-assembled JJ, in which two or more weakly coupled superfluids are spontaneously formed by the periodic density modulation of a supersolid~\cite{Wenzel2017,Biagioni2024}. In this system, the Josephson effect provides an opportunity to directly estimate the superfluid fraction. This calls for a better understanding of the dynamics of the Josephson oscillations in the context of quantum liquids and the role played by the BMF corrections. 

In this work, we consider a simple scenario of a junction based on a one-dimensional ultracold dilute quantum liquid formed by an unpolarized binary Bose mixture, where two bosonic species are close to the balance of attractive and repulsive contact interactions. In this regime, a liquid droplet would emerge without the presence of a potential barrier. An analogous three-dimensional system has been recently studied in Ref.~\cite{Liu2024} within the TMM approximation. Here we focus on understanding how the BMF effects affect the nature of the small-amplitude oscillations dynamics and the frequency of the oscillations in the Josephson regime; such shift in the frequencies would be a clear and experimentally measurable sign of the BMF character of the interactions. 

This work is organized as follows. In Sec.~\ref{sec.model}, we present the mean-field model of a 1D quantum liquid and discuss the pertinent approximations, introduce the double-well external potential defining the JJ, and arrive at the extended GPE which we subsequently solve in different interaction regimes. In Sec.~\ref{sec.tmm}, we introduce the TMM as the simplest approximation of the Josephson dynamics and derive the ordinary differential equations of the dynamical system governing the two-mode dynamics for the case of our extended GPE. In Sec.~\ref{sec.bogol}, we introduce the Bogoliubov approximation for the excitations in our model and, in Sec.~\ref{sec.res}, we present and discuss the numerical results in the repulsive and attractive mean-field regime, respectively. Finally, we discuss and summarize the results in Sec.~\ref{sec.conc}. 

\section{Model and methods}
\label{sec.model}
We study a 1D binary Bose mixture in a double-well potential
\begin{equation}\label{eq.potential}
V(x) = \frac12m \omega^2_{\rm ho} x^2 + V_0 e^{-2 x^2 / w^2} ~,
\end{equation}
where $\omega_{\rm ho}$ is the frequency of the harmonic trap and $V_0$ is the height of the potential barrier; see Fig.~\ref{fig.wavef}. We describe the two fully coherent condensates by mean-field condensate wave functions $\psi_\sigma(x, t)$ with the two species labeled as $\sigma = \uparrow, \downarrow$. The densities are given by $n_\sigma=|\psi_\sigma|^2$ and the coupling constants are denoted as $g_{\sigma \sigma}$ and $g_{\uparrow \downarrow}$ for the intraspecies and interspecies interactions respectively. In order to include the BMF correction, we follow the procedure outlined in Ref.~\cite{Tylutki2020}.

We assume that the mixture always remains in the miscible phase. It is further assumed that the residual mean-field interaction $\delta g = g_{\1\2}+\sqrt{g_{\1\1}g_{\2\2}}$ is weak. 
In the regime of low-energy excitations, where $n_\1 / n_\2 = \sqrt{g_{\2\2} / g_{\1\1}}$, the system is described by a single GPE for the mixture, which takes a dimensionless form~\cite{Tylutki2020},
\begin{align}\label{eq.gpe}
i \partial_t {\Psi} = \left( -\frac12\partial_{x}^2 + V(x) + s |\Psi|^2 - |\Psi| \right) \Psi ~,
\end{align}
where we used the healing length as a unit of length $\xi = \pi \hbar^2 \sqrt{2 |\delta g|} / [m \sqrt{g_{\1\1}g_{\2\2}} (\sqrt{g_{\1\1}} + \sqrt{g_{\2\2}})]$ and $\tau = m \xi^2 / \hbar$ as a unit of time. The sign of the residual interaction is denoted as $s = {\rm sgn} (\delta g)$. The dimensionless form for the external potential is $V(x) = \frac12 \omega_{\rm ho} x^2 + V_0 e^{-2 x^2/ w^2}$, while the frequencies are expressed in units of $1 / \tau$, the potential barrier in units of $m \xi^2 / \hbar^2$, and the condensate wave function $\Psi$ in units of $(\sqrt{g_{\1\1}} + \sqrt{g_{\2\2}})^{3/2} / [\sqrt{\pi\xi} (2 |\delta g|)^{3/4}]$. The rescaled normalization plays the role of a dimensionless control parameter, which depends on the interactions and the number of particles, 
\begin{equation}
N = \int |\Psi(x)|^2 dx ~.
\end{equation}
For large values of $N$ the mean-field term dominates over the LHY correction and Eq.~(\ref{eq.gpe}) reduces to the standard GPE, 
\begin{align}\label{eq.gpe0}
i \partial_t {\Psi} = \left( -\frac12\partial_{x}^2 + V(x) + s |\Psi|^2 \right) \Psi ~,
\end{align}
with the same normalization condition, which parametrizes the interaction strength. We therefore compare the solutions of Eq.~(\ref{eq.gpe}) with the dynamics of a standard BEC, described by Eq.~(\ref{eq.gpe0}). We show the example of the condensate ground states for the cases with and without the LHY term in Fig.~\ref{fig.wavef}. 

We use characteristic values of the parameters, $\omega_{\rm ho} = 1$ and $w = 0.5$, for solving Eqs.~(\ref{eq.gpe}) and~(\ref{eq.gpe0}), similarly to parameters commonly used to describe the standard JJ (see, for example, Ref.~\cite{Burchianti2017}). The initial configuration for the subsequent evolution in time is obtained as the ground state of the system either with a small offset for the barrier's position or by adding a slight linear tilt to the potential (we check our results with both methods) in order to introduce the initial imbalance $z_0 = z(0)$. The imbalance between the two wells is defined as
\begin{eqnarray}
z(t) = \frac{N_L(t) - N_R(t)}{N_L(t) + N_R(t)} ~,
\label{eq.imbalance}
\end{eqnarray}
and the number of particles in the left (right) well is calculated as $N_L(t) = \int_{-\infty}^0 |\Psi(x, t)|^2 dx$, ($N_R(t) = \int_{0}^\infty |\Psi(x, t)|^2 dx$). We also calculate the relative phase defined as
\begin{eqnarray}
\varphi(t) = \varphi_L(t) - \varphi_R(t) \quad \mod 2 \pi ~,
\label{eq.relph}
\end{eqnarray}
where the phases in the left (right) well are measured at the center of the corresponding well. We vary the imbalance in the entire possible range in order to scan across all dynamical regimes, from small plasma oscillations to self-trapping phenomena. The relation between the potential tilt (or the offset) and the imbalance is highly nonlinear and does not allow for a control of $z_0$ with arbitrary precision. Nevertheless, we can choose the values of $z_0$ from all dynamical regimes. However, in this article, we specifically focus on the dynamics of small-amplitude oscillations. The initial state is obtained from the imaginary-time evolution of Eq.~(\ref{eq.gpe}). Afterwards, the perturbation is removed and the real-time evolution is studied. 
\begin{figure*}
\includegraphics[clip = true, width = .9\textwidth]{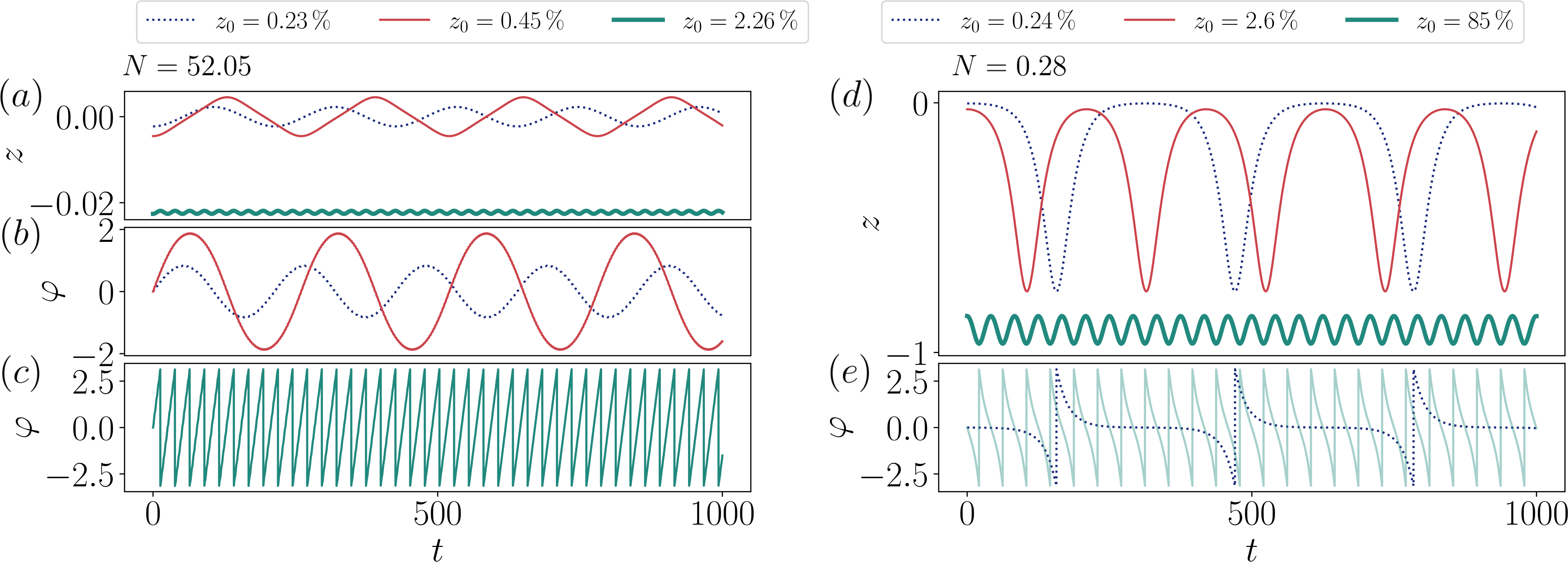}
\caption{Time-dependent GPE simulations in the repulsive regime, $\delta g > 0$, with the LHY correction ($V_0 = 10 \mu$ in all plots). The first column reports the results for a large normalization, $N = 52.05$. (a) $z(t)$ for two cases of Josephson oscillations (blue dotted and red solid lines) and self-trapping (thick cyan line). (b) $\varphi(t)$ in the Josephson regime and (c) $\varphi(t)$ for the self-trapped dynamics; same colors apply. Here, for small initial imbalance, we observe the regular plasma oscillations and the self-trapped regime is easily accessible when the imbalance grows above a certain threshold. The second column shows (d) the imbalance $z(t)$ and (e) relative phase $\varphi(t)$ for small normalization, $N = 0.28$. The localization-revival dynamics is observed for small initial imbalance $z_0$ (red solid and dotted blue lines). For large $z_0$, the system enters the self-trapped regime (thick cyan line). The corresponding phase dynamics is shown with the same colors. The normalization of the two examples corresponds to representative data points from Figs.~\ref{fig.freq-rep}(a),~\ref{fig.freq-rep}(b),~and~\ref{fig.freq-rep}(d). The initial imbalance is chosen arbitrarily to represent different dynamical regimes.}
\label{fig.td}
\end{figure*}
\begin{figure}
\includegraphics[clip = true, width = .98\columnwidth]{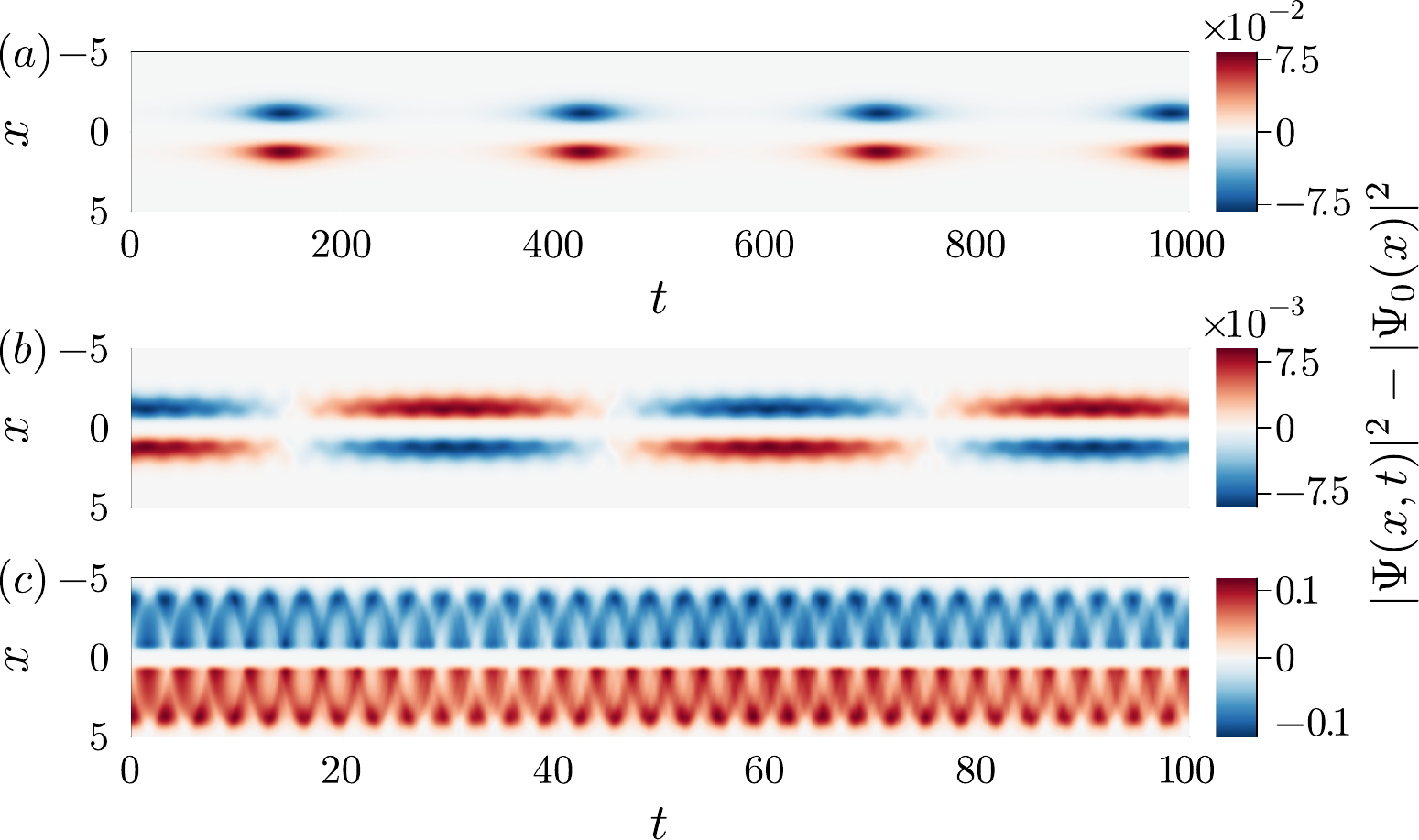}
\caption{Space-time plot showing the dynamics in three different regimes for the repulsive interactions, $\delta g > 0$, with the LHY term: (a) $N = 0.28$, (b) $N = 2.28$, (c) $N = 52.05$, and for $z_0 \approx 1\,\%$. The color scale represents the deviation from the equilibrium density during the time evolution, i.e., $|\Psi(x, t)|^2 - |\Psi_0(x)|^2$, where $\Psi_0$ is the lowest-energy symmetric solution of the GPE. }
\label{fig.cm}
\end{figure}
\section{Two-mode approximation}
\label{sec.tmm}
The simplest approach to model the dynamics of Josephson oscillations with the potential given by Eq.~(\ref{eq.potential}) is to use the celebrated two-mode approximation~\cite{Smerzi1997}, where the wave function is represented as
\begin{equation}\label{eq.tmmansatz}
\Psi(x,t) = \Phi_L(x)c_L(t)+\Phi_R(x)c_R(t)
\end{equation}
with $\Phi_{L/R}$ functions corresponding to the modes localized in the left and right well respectively and the complex coefficients $c_i(t)$ which in general vary with time. We determine the mode functions numerically as $\Phi_{L/R} = (\Psi_0 \pm \Psi_1) / \sqrt2$, where $\Psi_0$, ($\Psi_1$) are the lowest symmetric (antisymmetric) stationary solutions of the GPE~(\ref{eq.gpe}). The ansatz in Eq.~(\ref{eq.tmmansatz}) leads to 
\begin{equation}
i\partial_t c_L = E_0 c_L - J c_R + sU |c_L|^2 c_L - \alpha |c_L| c_L ~,
\label{eq:c_L}
\end{equation}
where
\begin{equation*}
J = -\int dx\, \Phi_L(x) \left( -\frac 12 {\partial_x^2} + V(x) \right) \Phi_R(x)
\end{equation*}
is the tunneling coefficient, 
\begin{equation*}
E_0 = -\int dx\, \Phi_L(x) \left( -\frac 12 {\partial_x^2} + V(x) \right) \Phi_L(x)
\end{equation*}
is the on-site energy, $U = \int{dx\, |\Phi_L(x)|^4}$ describes the interactions, $\alpha = \int dx\, |\Phi_L(x)|^3$ stems from the beyond-mean-field correction, and we neglected all interaction terms between different modes. The equation for $c_R$ is analogous to Eq.~(\ref{eq:c_L}). We express $c_i$ in terms of its phase and absolute value $c_i(t) = \sqrt{N_i(t)} \exp(i\varphi_i(t))$ and calculate the imbalance $z(t)$ as in Eq.~(\ref{eq.imbalance}) and the relative phase $\varphi(t)$ as in Eq.~(\ref{eq.relph}). Linearization of the resulting equations, i.e. assuming $z \ll 1$ and $\varphi \ll 1$, leads to
\begin{subequations}
\begin{eqnarray}
\partial_t z & = & 2J \varphi(t)\\
\partial_t \varphi & = & -\left( 2J + sNU - \alpha \sqrt{N / 2} \right) z(t) ~.
\end{eqnarray}
\end{subequations}
The imbalance $z(t)$, given by Eq.~(\ref{eq.imbalance}), obeys the harmonic oscillator equation, $\ddot{z}(t) + \omega_J^2 z(t) = 0$, and should therefore perform harmonic oscillations with the Josephson frequency, which yields
\begin{equation}\label{eq.omegaJ-tmm-lhy} 
\omega_{J} = \sqrt{2J \left( 2J + sNU - \alpha \sqrt{N / 2} \right)} ~.  
\end{equation}
The above result reduces to the well-known formula
\begin{equation}\label{eq.omegaJ-tmm} 
\omega_{J} = \sqrt{2J \left( 2J + sNU \right)} ~, 
\end{equation}
when the LHY correction is absent. The immediate observation is that the presence of the LHY correction makes the Josephson frequency in Eq.~(\ref{eq.omegaJ-tmm-lhy}) slightly smaller than the mean-field prediction given in Eq.~(\ref{eq.omegaJ-tmm}), provided the same barrier height is used. Indeed, it is known that for the Josephson junction in the mean-field BEC, the threshold for observing the self-trapped regime $z_c$ decreases with $N$ and is given by $z_c = 2 \sqrt{2 J N U - 4 J^2} / (N U)$~\cite{MartinezGaraot2018}. 

\section{Bogoliubov excitations}
\label{sec.bogol}
A more precise estimation of the Josephson frequency can be obtained through the analysis of the Bogoliubov excitation spectrum. We consider a small perturbation $\delta \Psi(x, t)$ of the symmetric (i.e., unpolarized) lowest-energy state $\Psi_0(x)$ and expand the wave function as $\Psi(x, t) = e^{-i \mu t} \Psi_0(x) + \delta \Psi(x, t)$ up to linear terms in $\delta \Psi$. 

Then, expanding the perturbation into modes as
\begin{equation}
\delta \Psi(x, t) = \sum_\eta \left( u_\eta(x) e^{-i (\mu + \omega_\eta) t} + v^*_\eta(x) e^{-i (\mu - \omega_\eta) t} \right)
\end{equation}
leads to the Bogoliubov--de Gennes (BdG) equations in the form
\begin{equation}
\begin{pmatrix}
K - \omega_\eta & L \\
L & K + \omega_\eta \\
\end{pmatrix} \begin{pmatrix} u(x)\\ v(x)\\ \end{pmatrix} = 0 ~,
\end{equation}
where $u_\eta(x)$ and $v_\eta(x)$ are the wave functions of the Bogoliubov modes. The operators in the BdG matrix take the form 
\begin{subequations}
\begin{align}
K &= -\frac12 \partial^2_x - \mu + 2 s \Psi_0^2 - \frac32 \Psi_0\\
L &= s \Psi_0^2 - \frac12 \Psi_0\\ \nonumber
\end{align}
\end{subequations}
when the LHY term is included~\cite{Tylutki2020}. When the LHY correction is absent, these expressions simplify as $K = -\partial^2_x / 2 - \mu + 2 s \Psi_0^2$ and $L = s \Psi_0^2$. 

The Josephson frequency $\omega_J$ is given by the first nonzero excitation frequency above the lowest energy mode, i.e., $\omega_J = \omega_1$, as the first antisymmetric mode corresponds to the oscillatory motion between the two wells. It is well known that the Bogoliubov spectrum gives a better estimation of the Josephson frequency than the TMM~\cite{Burchianti2017}. The appearance of imaginary frequencies in the spectrum is a hallmark of an instability of the underlying state around which the expansion is performed and the presence of a true ground state with a lower energy. In this case, the dynamics is analyzed with respect to the false ground state. 

\section{Results}\label{sec.res}
\subsection{Repulsive Interactions}
\begin{figure*}[!t]
\includegraphics[clip = true, width = .99\textwidth]{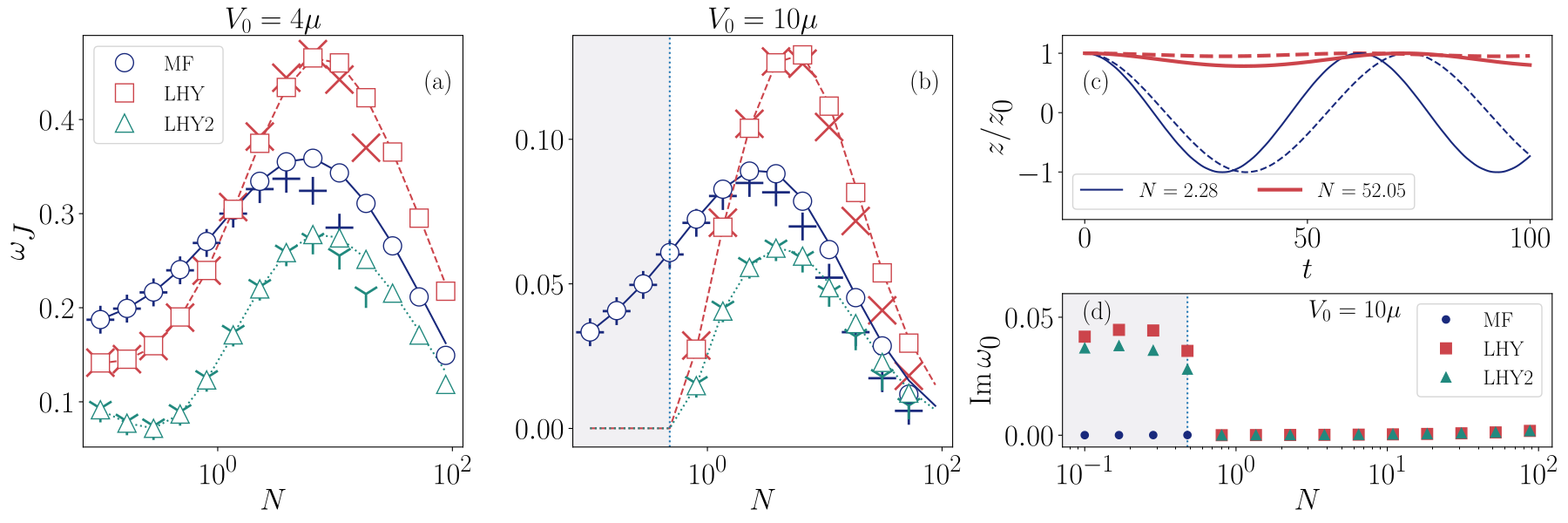}
\caption{Josephson frequencies for the overall repulsive mean-field interaction $\delta g > 0$ for two barrier heights: (a) $V_0 = 4 \mu$ and (b) $V_0 = 10 \mu$. In both panels, we show the solutions for the pure mean-field case (MF), given by Eq.~(\ref{eq.gpe0}), the case with the LHY correction (LHY), given by Eq.~(\ref{eq.gpe}), and the case with the LHY correction but with the barrier that is the same as for the mean-field case (LHY2), in the units of Eq.~(\ref{eq.gpe}). Empty markers correspond to fits to the solutions of time-dependent GPE~(\ref{eq.gpe}): blue circles (MF), red squares (LHY), and cyan triangles (LHY2). Lines correspond to the Bogoliubov frequencies: blue solid (MF), red dashed (LHY), and cyan dotted (LHY2) respectively. The results of the TMM, that is, with the LHY term, given by Eq.~(\ref{eq.omegaJ-tmm-lhy}), and without the LHY term, given by Eq.~(\ref{eq.omegaJ-tmm}), are shown with blue pluses ($+$), red crosses ($\times$) and cyan $\Ydown$'s, respectively, for MF, LHY, and LHY2. Note that in the regime of localization-revival dynamics, marked as a grey-shaded area, the Bogoliubov spectrum gives $\omega_J = 0$. (c) Two generic examples of time evolution of the imbalance rescaled to its initial value: $z(t) / z_0$ for the Josephson (blue) and self-trapped (red thick line) regimes, both with (solid lines) and without (dashed lines) the BMF interaction. The simulation parameters correspond to those from (b), that is, $V_0 = 10 \mu$. (d) The imaginary part of the first (i.e., symmetric) excitation frequency corresponding to $V_0 = 10 \mu$ that becomes purely imaginary in the regime of localization-revival dynamics for the quantum liquid (LHY and LHY2). The border between the Josephson and localization-revival (shaded area) regimes is shown as a dotted vertical line. When the BMF interaction is absent, the frequencies are real across the entire interaction regime (MF). }
\label{fig.freq-rep}
\end{figure*}

We first analyze the dynamics of Eq.~(\ref{eq.gpe}) with repulsive overall mean-field interactions, i.e. $\delta g > 0$, which is the regime where the Josephson dynamics is typically considered, and we study the role played by the BMF fluctuations. In the standard JJ, for a small initial imbalance $z_0$ the system performs plasma oscillations, i.e., the imbalance oscillates symmetrically around zero, and the phase difference between the wells shows the same behavior. This regime is well understood in terms of small-amplitude oscillations, as is clear from the Bogoliubov analysis (see, also, subsequent paragraphs). When the initial $z_0$ grows above a certain threshold, the system enters the dynamical regime of highly nonlinear oscillations, where the average occupation of one well is larger than the occupation of the other, and the system performs oscillations around this average imbalance. This regime is known as a self-trapping; a characteristic feature of this regime is that the relative phase $\varphi(t)$ grows linearly with time. 

The inclusion of the BMF correction shows that for intermediate to large (rescaled) particle number, the system follows the behavior described above with only quantitative corrections. We plot the time dependence of $z(t)$ in Fig.~\ref{fig.td}(a) for two cases in the Josephson regime (dotted blue and solid red lines) and one in the self-trapping regime (solid cyan line). The dynamics of the relative phase is shown in Figs.~\ref{fig.td}(b) and (c) with the same colors as in Fig.~\ref{fig.td}(a). The choice of normalization in these examples, i.e., $N = 0.28$ and $N = 52.05$, is consistent with the data presented in subsequent plots, i.e., Figs.~\ref{fig.freq-rep}(a), (b), (d), where we vary the normalization uniformly on a logarithmic scale. We also present the time evolution of the entire density distribution by plotting the deviation from the unpolarized density, $|\Psi(x, t)|^2 - |\Psi_0(x)|^2$, where $\Psi_0(x) = \Psi_0(-x)$ is the lowest-energy symmetric wave function, in Figs.~\ref{fig.cm}(b) and~\ref{fig.cm}(c) respectively, for the Josephson dynamics for intermediate $N$ and for the self-trapped dynamics for large $N$. The retrieval of these two regimes, present in the standard bosonic JJ, is explained by the large $N$, which leads to a strong mean-field interaction term that dominates over the BMF correction.

The difference between the dynamics of a quantum liquid and a mean-field condensate becomes evident for small $N$, where, instead, it is the LHY term that dominates over the mean-field interaction. In this regime, for small initial imbalance, we observe a completely different dynamics, which consists of a periodic recurrence of large-amplitude imbalance between the two wells intermitted with intervals of nearly symmetric occupations. Note that the revivals take place on a different timescale as compared to the regular Josephson oscillations. This localization-revival dynamics resembles the dynamics of the Jaynes-Cummings model. Using the two-mode approximation and expanding the wave function into components with varying population imbalance, i.e. $|n,N-n\rangle$ with $n$ bosons in the left and $N-n$ in the right well, would indeed lead to similar time evolution equations with revivals stemming from the interference of different occupation numbers. Similar dynamics has been studied for a weakly \emph{attractive} BEC in a double well-potential~\cite{Pawlowski2011}. Note that here the LHY attraction indeed overcomes the weakly repulsive mean-field term, which makes a formation of a bright soliton possible. We plot the time dependence of the imbalance, $z(t)$, for a trap with the barrier height of $V_0=10\mu$ in Fig.~\ref{fig.td}(d) for three different $z_0$, and the corresponding phase difference in Fig.~\ref{fig.td}(e). It becomes clear that when the initial imbalance is small, its subsequent dynamics undergoes periodic localization in one of the wells. However, when the initial imbalance exceeds a certain threshold, it enters the self-trapped regime with visibly different dynamics. To clearly visualize the time evolution of the entire density distribution, we plot the deviation of the density from its symmetric equilibrium in color in Fig.~\ref{fig.cm}(a). One can clearly see the periodic but nonlinear dynamics of localization revivals. 

Next, we focus on the Josephson regime studying the role of the BMF corrections on the Josephson plasma oscillations. To this end, we extract the frequencies of Josephson oscillations from our time-dependent simulations of Eq.~(\ref{eq.gpe}) and compare the results obtained both with and without the LHY term. With the growing normalization, the system becomes ever less sensitive to the presence of the barrier because the chemical potential also grows. Since the latter sets the energy scale for the system, in order to make our comparison comprehensive and reliably compare the results with different $N$, which we vary in a wide range between $N = 10^{-1}$ and $N = 10^2$, we decide to first measure the barrier height against the chemical potential. Then, we also perform the third series of simulations, for which we evolve Eq.~(\ref{eq.gpe}), but we fix the barrier height to that from the simulation of Eq.~(\ref{eq.gpe0}) with the same $N$. These results are shown in Fig.~\ref{fig.freq-rep}(a) for a relatively low barrier, $V_0 = 4 \mu$, and in Fig.~\ref{fig.freq-rep}(b) for a larger barrier, $V_0 = 10 \mu$. 

For low barrier heights, in our example for $V_0=4\mu$, we observe Josephson oscillations across the entire range of particle number $N$, regardless of whether we deal with a quantum liquid or a regular BEC. In Fig.~\ref{fig.freq-rep}(a), we show the results of the evolution of the GPE for a liquid, given by Eq.~(\ref{eq.gpe}), with red squares (labeled as LHY) and the standard mean-field GPE, given by Eq.~(\ref{eq.gpe0}), with blue circles (labeled as MF), and the result of Eq.~(\ref{eq.gpe}) but for the same barrier height that we had for Eq.~(\ref{eq.gpe0}) with cyan triangles (labeled as LHY2). It becomes clear that the presence of the LHY term, although moderately manifested in the density profiles (see Fig.~\ref{fig.wavef} for the wave-function profiles), significantly modifies the Josephson frequencies. For the sake of completeness, we also calculate the Josephson frequencies from the TMM with the LHY term, given by Eq.~(\ref{eq.omegaJ-tmm-lhy}), and without the LHY term, given by Eq.~(\ref{eq.omegaJ-tmm}), and we show the results with markers: blue pluses ($+$), red crosses ($\times$) and cyan Y's ($\Ydown$) for the same three cases respectively. Finally, we compare the results with the calculation of the Bogoliubov spectrum, which we mark with blue solid, red dashed, and cyan dotted lines, respectively, for the three cases. It is clear that for small to intermediate $N$, all three methods give the same frequencies, and the discrepancies between the TMM and the two other methods appear for large $N$, where the nonlinearities start playing an important role and in agreement with the observations of Ref.~\cite{Burchianti2017} for a regular GPE. 

For larger barriers, such as $V_0 = 10\mu$ in our example, the effect of the LHY term becomes more pronounced and might result in a qualitatively different behavior. For intermediate and large $N$, we observe regular Josephson dynamics as we did for the lower barrier, and all three methods of calculating the Josephson frequencies give similar results [see, also, Fig.~\ref{fig.freq-rep}(c) for $z(t)$ as a function of time]. However, the difference becomes dramatic in the small interaction regime, $N \ll 1$, where the presence of the LHY correction has a profound effect on the oscillation dynamics. In that case, the system enters the localization-revival regime shown as a grey area in Fig.~\ref{fig.freq-rep}(b). It is remarkable that in this regime, the Bogoliubov analysis gives a zero frequency corresponding to the first antisymmetric mode (oscillations); see the red dashed and cyan dotted lines in Fig.~\ref{fig.freq-rep}(b). Moreover, the frequency of the lowest mode acquires a finite value and becomes imaginary, clearly showing the instability of this state, as shown in Fig.~\ref{fig.freq-rep}(d). Note that this is another striking difference between the nature of the localization-revival dynamics and the self-trapping. 
\begin{figure}
\includegraphics[clip = true, width = .99\columnwidth]{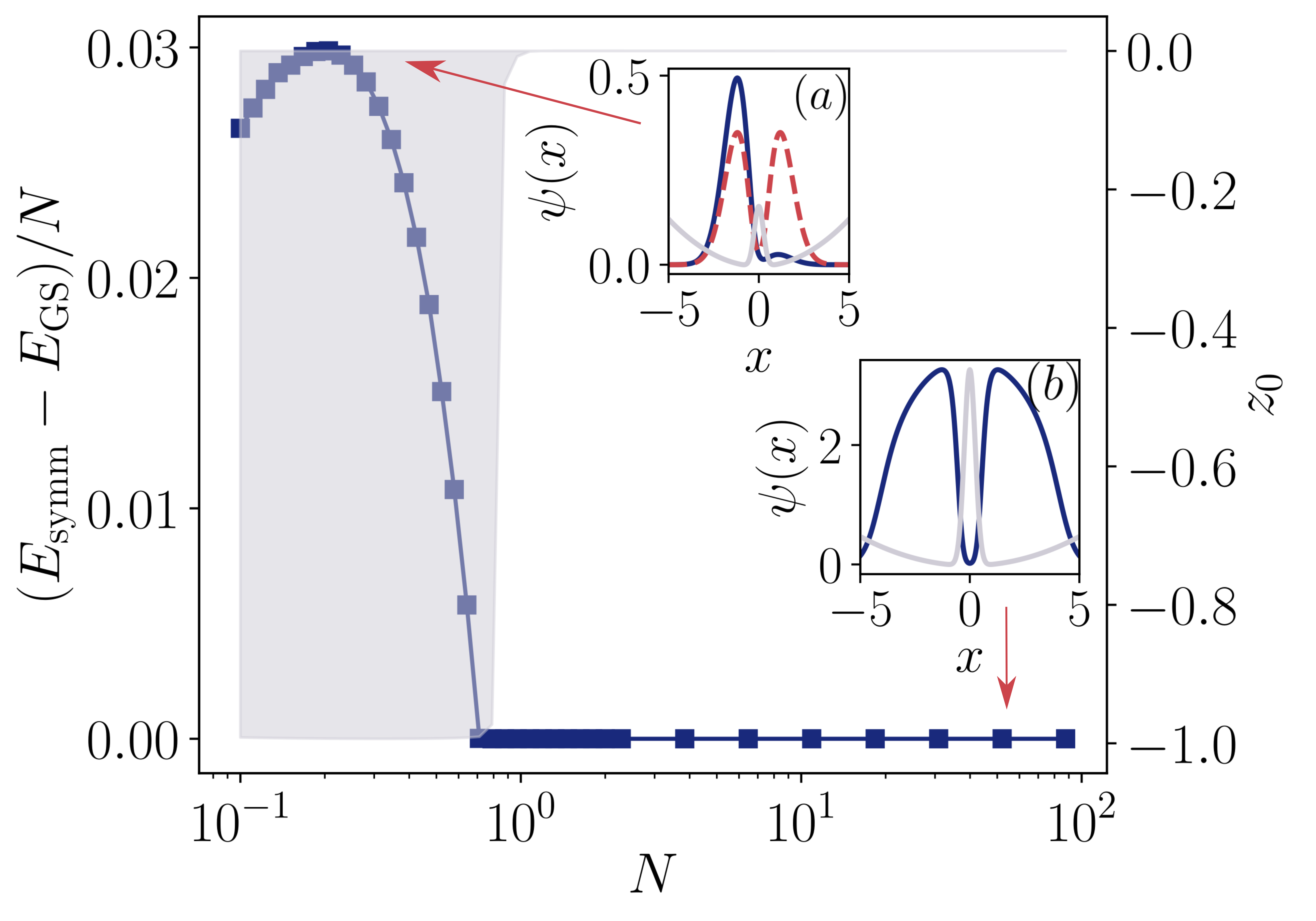}
\caption{The difference between the energy of an unpolarized solution and the ground-state energy per particle, $(E_{\rm symm} - E_{\rm GS}) / N$, for the same parameters as in Fig.~\ref{fig.freq-rep}(b), that is, $V_0 = 10 \mu$ and $\delta g > 0$ (blue squares and line). The corresponding polarization (imbalance) of the ground state is shown as a gray shade with a scale on the right axis. Below the critical $N$, the ground state becomes highly polarized (localization) and the unpolarized state is unstable. These results remain in agreement with the Bogoliubov spectrum and time dynamics. The insets show (a) the wave functions of the polarized ground state (blue solid line) and unpolarized unstable state for the localization regime and (b) the unpolarized ground state (blue solid line) for the Josephson regime. For clarity, the shape of the potential is also shown (grey line, not in scale).}
\label{fig.gsener}
\end{figure}
Such behavior can be further explained by studying how the actual ground state changes when $N$ decreases below this threshold value. It turns out that in agreement with the Bogoliubov analysis the symmetric, unpolarized solution no longer corresponds to a stable energy minimum, but the true ground state becomes polarized, as shown in Fig.~\ref{fig.gsener}. The energy of the unpolarized state is larger than the ground-state energy exactly below the same value of $N$ for which we observe the transition to the localization-revival dynamics. The ground states are shown in the insets of Fig.~\ref{fig.gsener}(a) for the polarized phase (small $N$) and Fig.~\ref{fig.gsener}(b) for the unpolarized phase (large $N$). 

\subsection{Attractive Interactions}
\begin{figure}
\includegraphics[clip = true, width = .99\columnwidth]{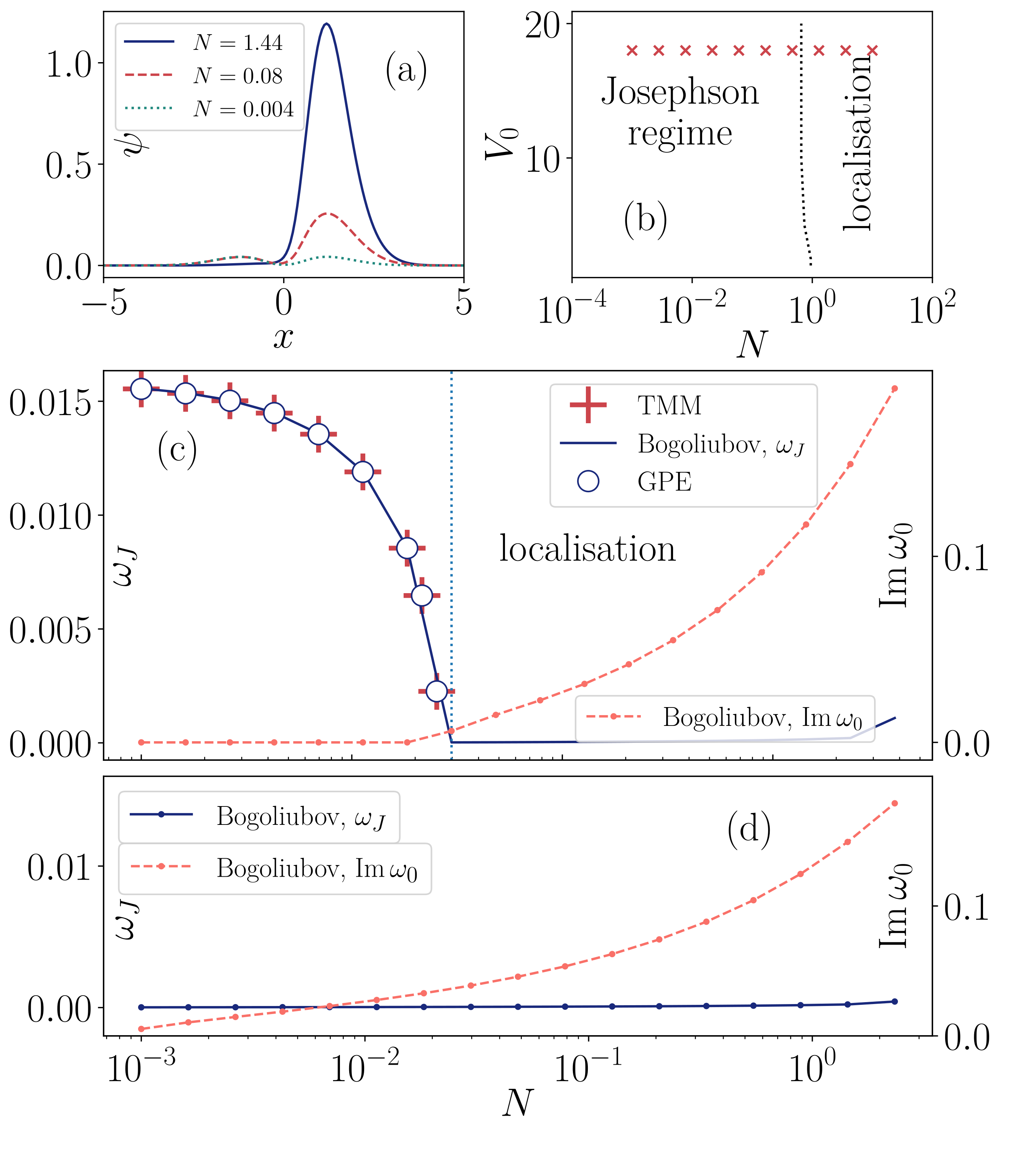}
\caption{Results for the overall attractive mean-field interaction, $\delta g < 0$. (a) The wave functions for three different $N$ for the case of the standard mean-field GPE~(\ref{eq.gpe0}) without the LHY term. (b) Phase diagram corresponding to GPE without the LHY terms, given by Eq.~(\ref{eq.gpe0}), in the small-amplitude oscillations regime, i.e., the Josephson dynamics vs the localization-revival dynamics. Crosses correspond to the parameters we chose for (c) and (d). (c) Calculated frequencies of the Josephson oscillations of Eq.~(\ref{eq.gpe0}) as a function of $N$: GPE (circles), TMM (crosses), and Bogoliubov frequency (lines). The scale is on the left. The orange dashed line shows the imaginary part of the lowest (symmetric) mode frequency, where a nonzero value signifies the presence of the instability (scale on the right). 
(d) Same as in (c), but for the case with the BMF correction, i.e., Eq.~(\ref{eq.gpe}). }
\label{fig.freq-att}
\end{figure}
Finally, we analyze the dynamics of the oscillations for the attractive mean-field interaction, i.e., $\delta g < 0$. In contrast to the procedure used in the repulsive case studied in the previous section, we maintain a constant barrier height in units of Eq.~(\ref{eq.gpe}), rather than keeping the $V_0 / \mu$ ratio fixed (but in a wide range corresponding to $V_0 / \mu \approx 10$), as the chemical potential changes very rapidly in the attractive case. In the absence of the BMF correction and for sufficiently low particle numbers, the dynamics of the system is characterized by low-frequency Josephson oscillations. Again, we calculate the Josephson frequencies solving the time-dependent GPE~(\ref{eq.gpe0}) and from the TMM, given by Eq.~(\ref{eq.omegaJ-tmm}), and, finally, as the first antisymmetric Bogoliubov excitation. We find a very good agreement between all three approaches and the results are presented in Fig.~\ref{fig.freq-att}(c) with a blue line for the Bogoliubov frequency, circles for the GPE solution, and red crosses for the TMM (the scale is given by the vertical axis on the left). We also plot the imaginary part of the frequency of the lowest symmetric mode frequency with a dashed orange line (scale on the right). Since in that case the mean-field interaction is attractive, the transition to localization occurs even in the absence of the BMF correction~\cite{Trenkwalder2016}. The mechanism resembles that of the repulsive mean-field case: above a certain threshold, the attraction becomes sufficiently strong to destabilize the unpolarized state towards the true ground state, which becomes polarized [see Fig.~\ref{fig.freq-att}(a) for the wave-function profiles]. This effect does not strongly depend on $V_0$, as shown in the phase diagram in Fig.~\ref{fig.freq-att}(b). 

The presence of the LHY term makes the ultradilute quantum liquid with attractive interactions always localize, regardless of the interaction regime. In this case, the LHY correction is strong enough to compensate for the weak mean-field attraction in the small-$N$ regime. The results of the Bogoliubov spectrum calculation are depicted in Fig~\ref{fig.freq-att}(d). As shown, the Josephson frequency remains zero across the entire range of $N$ (solid blue line, scale on the left), while the imaginary part of the frequency of the lowest symmetric mode is nonzero everywhere (dashed orange line, scale on the right). In the regime of small interactions, where the LHY term dominates, the system asymptotically approaches the pure LHY liquid obtained in Ref.~\cite{Skov2021}; therefore, the JJ under these conditions can be another way to test the LHY quantum liquid experimentally, and the physics governed by corrections beyond the LHY term. 

\section{Conclusions}
\label{sec.conc}
In conclusion, we analyzed the small-amplitude oscillations of the one-dimensional bosonic JJ for an ultradilute quantum liquid and compared the results with that of the usual BEC governed by the GPE. This allowed us to investigate the role of the BMF corrections on the Josephson dynamics. The results indicate that apart from the regular Josephson oscillations there is a highly nonlinear regime corresponding to localization-revival dynamics. This regime emerges in quantum liquids either when the overall mean-field interaction is repulsive for weak interactions or when it is attractive for any interaction strength.

When an ultradilute liquid stays in the regime of regular Josephson oscillations, its frequencies significantly differ from those of a regular BEC, providing, in principle, yet another way to characterize the ultradilute liquids in the experiment. Note that our model is strictly one dimensional and its direct experimental realization is challenging. In fact, recent experimental realizations of binary quantum droplets~\cite{Cabrera2018, Cheiney2018} and pure LHY quantum liquid~\cite{Skov2021} remained in the confined three-dimensional regime where local density approximation works well, resulting in a different type of nonlinearity than we consider here~\cite{Debnath2020, Debnath2021}. Studying Josephson dynamics in quasi-1D ultradilute liquids across the dimensional crossover~\cite{Ilg2018,Zin2018} would constitute a valuable extension of this work, which we leave as an outlook for the future. 

Our study, despite being limited to one dimension, sets the stage for more complex scenarios, such as the aforementioned quasi-1D geometries or those involving dipolar gases, where a self-organized JJ was recently observed~\cite{Biagioni2024}. 

\bigskip \noindent
\section*{Acknowledgments}
This work was supported by the National Science Centre, Poland (NCN), Contracts No. UMO-2019/35/D/ST2/00201 (P.W. and M.T.) and No. 2020/37/B/ST2/00486 (K.J.). G.E.A. acknowledges financial support from Ministerio de Ciencia e Innovación MCIN/AEI/10.13039/501100011033 (Spain) under Grant No.~PID2020-113565GB-C21 and from AGAUR-Generalitat de Catalunya Grant No.~2021-SGR-01411.

%

\end{document}